\documentclass[12pt,aps,showpacs]{revtex4}
\usepackage{amssymb}
\usepackage[dvips]{graphicx}
\usepackage{amsmath}
\usepackage{amssymb}
\usepackage{amsmath,amssymb,graphics,epsfig,color,times,bbm}
\usepackage{caption}
\usepackage{soul}

\soulregister\Hl{7}
\soulregister\ref7
\soulregister\cite7

\begin{document}

\author{vskip }
\affiliation{vskip }

\captionsetup[figure]{labelfont={bf},name={Fig.},labelsep=period,singlelinecheck=off,justification=raggedright}

\bigskip \noindent \textbf{\large High-performance cavity-enhanced quantum
memory with warm atomic cell}

\noindent {Lixia Ma$^{1,3}$, Xing Lei$^{1,3}$, Jieli Yan$^{1}$, Ruiyang Li$%
^{1}$, Ting Chai$^{1}$, Zhihui Yan$^{1,2\ast }$, Xiaojun Jia$^{1,2\dagger }$%
, Changde Xie$^{1,2}$ \& Kunchi Peng$^{1,2}$}

\noindent {\small $^{1}$State Key Laboratory of Quantum Optics and Quantum
Optics Devices, Institute of Opto-Electronics, Shanxi University, Taiyuan
030006, P. R. China\newline
$^{2}$Collaborative Innovation Center of Extreme Optics, Shanxi University,
Taiyuan 030006, P. R. China\newline
$^{3}$These authors contributed equally: Lixia Ma, Xing Lei.\newline
$^{\ast }$e-mail:zhyan@sxu.edu.cn\newline
$^{\dagger }$e-mail:jiaxj@sxu.edu.cn}


\section*{Abstract}

\vskip0.05truein \textbf{\noindent High-performance quantum memory for
quantized states of light is a prerequisite building block of quantum
information technology. Despite great progresses of optical quantum memories
based on interactions of light and atoms, physical features of these
memories still cannot satisfy requirements for applications in practical
quantum information systems, since all of them suffer from trade-off between
memory efficiency and excess noise. Here, we report a high-performance
cavity-enhanced electromagnetically-induced-transparency memory with warm
atomic cell in which a scheme of optimizing the spatial and temporal modes
based on the time-reversal approach is applied. The memory efficiency up to
67}$\pm $\textbf{1\%\ is directly measured and a noise level close to
quantum noise limit is simultaneously reached. It has been experimentally
demonstrated that the average fidelities for a set of input coherent states
with different phases and amplitudes within a Gaussian distribution have
exceeded the classical benchmark fidelities. Thus the realized quantum
memory platform has been capable of preserving quantized optical states, and
is ready to be applied in quantum information systems, such as distributed
quantum logic gates and quantum-enhanced atomic magnetometry.}

\section*{Introduction}

The high-performance quantum memory featuring both high memory efficiency
and low excess noise is an indispensable building block in quantum
information systems, distributed quantum computation and quantum metrology.
For example, the multiple spatial separated macroscopic objects can be
entangled by efficiently storing multipartite entangled state of optical
modes, and its realization depends on memory efficiency and noise level of
the multipartite entangled optical state \cite{1}. In distributed quantum
computation, enhancing the memory fidelity among entangled different modules
(nodes) is significant for implementing optical logical gates\textbf{\ }\cite%
{2,3}. It has been demonstrated that the atom-based measurement sensitivity
is ultimately restricted by the quantum noise limit (QNL), and spin
squeezing holds a promise to overcome this restriction \cite{4,5}. In
quantum-enhanced atomic magnetometry, the spin squeezing can be generated by
efficiently storing the squeezed optical mode, and then used to measure weak
signal merged in the quantum noise \cite{6}.

Over the past decades, various light-atom interactions have been utilized to
implement quantum memory, such as electromagnetically-induced-transparency
(EIT) \cite{7,8,9,10,11,12}, far-off-resonance Raman \cite{13}, quantum
non-demolition \cite{14}, Autler--Townes splitting \cite{15} and photon echo
interaction \cite{16}. By applying a gradient magnetic field in an
atomic cell operating with a special regime, gradient echo memory provides the storage of coherent pulses containing around one photon with
the recall fidelity up to 98\% \cite{17}. The high memory efficiency is
crucially important for practical quantum information \cite{18,19}. Towards
the aim of high efficiency, the coherent optical storage efficiency has
reached 92.0$\pm $1.5\%\ in an optically dense cold atomic media based on
EIT effect, however, the excess noise attached to the signal mode is about
6\% higher than the QNL \cite{20}. The memory efficiency in warm
atomic cell has also been improved by using optimal input signal mode \cite%
{21}. Besides, another challenge for the memory of quantized optical states
is to suppress the excess noise, which will destroy quantum features of
stored states. In the processing of atom-light interaction with a $\Lambda $%
-type energy configuration, the coupling of the control mode on the signal
mode transition will induce unwanted four-wave-mixing (FWM) noise, which is
the main noise source of quantum memory \cite{22,23}. In near-resonant EIT
memory with a moderate atomic number, the interaction for storing quantum
states is dominant and the influence of FWM noise is relatively less. The
EIT memory in the warm atomic vapor with the noise of 2\% higher than the
QNL and the completed storage amplitude efficiency of 10\% has been reported 
\cite{24}.\textbf{\ }It has been proved that EIT memories are able to reach
the QNL, which have been applied to preserve the squeezed light with atomic
ensemble \cite{25,26}. Alternatively, the optical cavity can enhance the
light-atom interaction \cite{27,28,29,30,31,32,33} and suppress the excess
noise \cite{34}. Although in a demonstrated cavity-enhanced Raman memory the
FWM noise has been effectively suppressed, the memory efficiency is below
10\% \cite{22}. So far, the effective realization of quantum memory
with both high efficiency and low noise approaching the QNL is still a
significant challenge for practical applications.

To accomplish quantum memory with the necessary features of both high memory
efficiency and low excess noise, we present an experimental
demonstration on a cavity-enhanced EIT memory in a simple warm atomic cell, in which the near-perfect mode
matching technique based on time-reversal approach is applied \cite{21}. In
this way, not only the FWM noise but also the
other noises are actively suppressed, due to off resonance with optical
cavity; while the memory interaction is effectively enhanced, by resonating
the signal mode in the optical cavity. Thanks to the cavity-enhanced EIT
interaction and the near-perfect mode matching, both high memory efficiency
up to 67$\pm $1\%\ and low noise level close to the QNL are simultaneously
obtained by the memory system. Based on both high efficiency and low excess
noise at QNL level, the average fidelities of the memory measured on a set
of input coherent states with varied phases and amplitudes within a Gaussian
distribution have totally exceeded the corresponding classical benchmark
fidelities. Thus, the performance of the memory has reached a level higher
than any classcial memory and is able to store quantum states of light \cite%
{35,36}.

\section*{Results}

\textbf{Principle of cavity-enhanced quantum memory.} In quantum optics, the
optical mode is represented by the annihilation operator $\hat{a}$, the
amplitude (phase) quadrature $\hat{X}_{L}$ $(\hat{Y}_{L})$ of light
corresponds to the real (imaginary) part of the annihilation operator $\hat{a%
}$, as $\hat{X}_{L}=(\hat{a}+\hat{a}^{\dag })/\sqrt{2}$ ( $\hat{Y}_{L}=(\hat{%
a}-\hat{a}^{\dag })/\sqrt{2}i$) \cite{37}. Under the Holstein-Primakoff
approximation, the collective atomic spin wave is described by the lowering
operator $\hat{S}=\sum_{i}\left\vert g\right\rangle \left\langle
m\right\vert $, here $|g\rangle $ and $|m\rangle $ stand for a ground state
and meta-stable state, respectively. The amplitude (phase) quadrature $\hat{X%
}_{A}$ $(\hat{Y}_{A})$ of the atoms is associated with the $y$ $(z)$
component of the Stokes operator $\hat{S}_{y}$ $(\hat{S}_{z})$ on the Bloch
sphere, which is represented by $\hat{X}_{A}=(\hat{S}+\hat{S}^{\dag })/\sqrt{%
2}=\hat{S}_{y}/\sqrt{\left\langle \hat{S}_{x}\right\rangle }$ $(\hat{Y}_{A}=(%
\hat{S}-\hat{S}^{\dag })/\sqrt{2}i=\hat{S}_{z}/\sqrt{\left\langle \hat{S}%
_{x}\right\rangle })$ \cite{14}. The coherent state of light is a minimum
uncertainty state with equal uncertainty between two conjugate quadrature
components, which is usually used to describe the quantized state of laser.
The quantum natures of an optical memory can be characterized by means of
preserving coherent state, thus the coherent state of optical mode is
utilized as the input state of quantum memory in our experiment. The
quantized state can be transferred between light and atomic superposition in
the EIT memory \cite{8}. The $\Lambda $-type three-level system of a ground
state $|g\rangle $, a meta-stable state $|m\rangle $ and an excited state $%
|e\rangle $ is employed in the EIT configuration, which is presented in the
insert of Fig. 1 (a). The signal mode is near resonant with the transition
between a ground state $|g\rangle $ and an excited state $|e\rangle $, while
control mode is near resonant with the transition between a meta-stable
state $|m\rangle $ and an excited state $|e\rangle $. In our system, the
control mode is much stronger than the signal mode, and is treated as a
classical mode. When the collective atomic spin wave $\hat{S}(t)$ interacts
with the signal mode $\hat{a}(t)$ via EIT process, the quantized state of
the signal mode and the atomic ensemble can be transferred to each other,
because the effective Hamiltonian $\hat{H}_{EIT}$ of light-atom interaction
is a type of beam-splitter interaction \cite{38}. The quantum memory process
includes three stages of writing, storage, and reading which are implemented
by modulating the light-atom interaction with a control mode. Therefore, the
step-like function used as an approximation of switching on and off
processes in EIT interaction can be shown as follows: $\hat{H}(t)=\hbar
\kappa \hat{a}^{\dag }\hat{S}+\hbar \kappa \hat{S}^{\dag }\hat{a}$ $(-\infty
<t<0)$; $\hat{H}(t)=0$ $(0<t<T_{0})$; $\hat{H}(t)=\hbar \kappa \hat{a}^{\dag
}\hat{S}+\hbar \kappa \hat{S}^{\dag }\hat{a}$ $(T_{0}<t<\infty )$, where $%
T_{0}$ is the storage time, $\kappa =\sqrt{N_{a}}\mu \Omega /\Delta $ is the
effective light-atom interaction constant, $N_{a}$ is the atomic number, $%
\mu $ is the light-atom coupling constant,\ $\Omega $ is the Rabi frequency
of the control mode, and $\Delta $ is the detuning between light and atom
coupling.

\begin{figure}[tbph]
\begin{center}
\includegraphics[width=16cm]{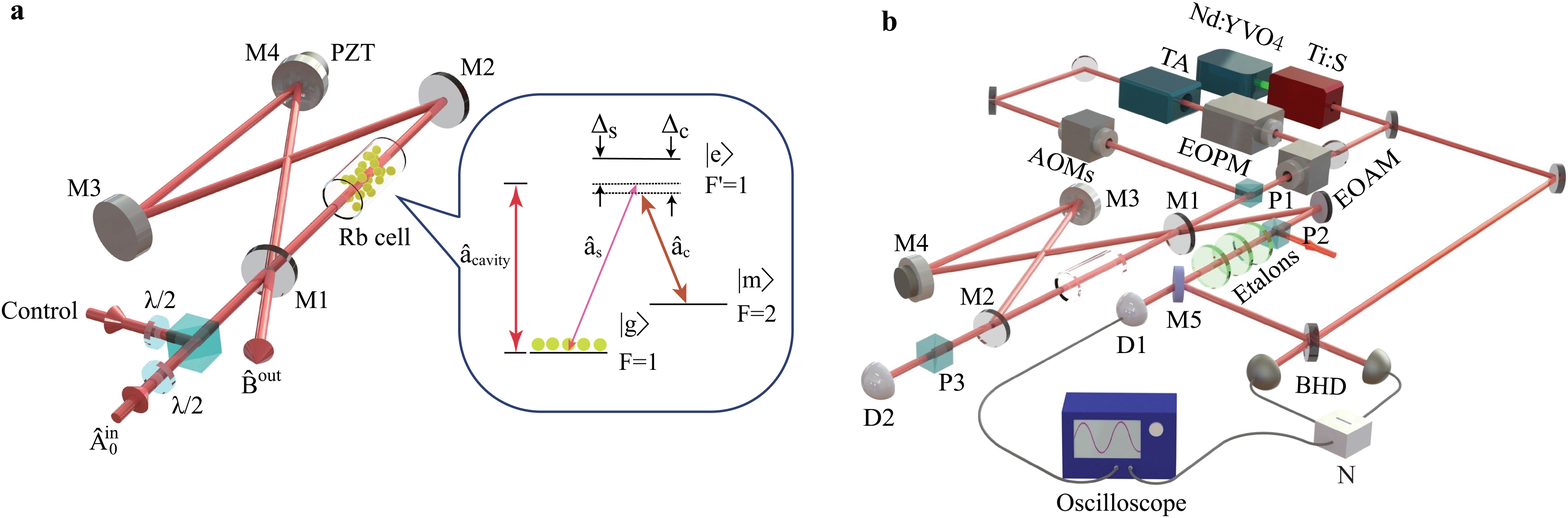}
\end{center}
\caption{\textbf{Schematic diagram.} \textbf{a} Diagram for the
cavity-enhanced quantum memory and atomic energy level for quantum memory.
Atoms with a ground state $\left\vert g\right\rangle $: $\left\vert
5S_{1/2},F=1\right\rangle $, a meta-stable state $\left\vert m\right\rangle $%
: $\left\vert 5S_{1/2},F=2\right\rangle $, and an excited state $\left\vert
e\right\rangle $: $\left\vert 5P_{1/2},F^{\prime }=1\right\rangle $ are
shown. The atomic cell is placed between two plano mirrors, where the
optical cavity simultaneously enhances the light-atom interaction and
suppresses the excess noise. \textbf{b} Experimental setup for implementing
cavity-enhanced quantum memory. TA: tapered amplifier; PZT, piezoelectric
transducer; EOAM, electro-optical amplitude modulator; EOPM, electro-optical
phase modulator; AOM, acousto-optical modulator; BHD, balanced homodyne
detector; D, photoreceiver detector; M, mirror; P, Glan-Thompson polarizer; N, negative power combiner.}
\end{figure}

Fig. 1 (a) is a diagram for the cavity-enhanced quantum memory with a warm
atomic cell. The cavity with a bow-tie-type ring configuration consists of
two plano mirrors and two concave mirrors, which enables to enhance the
light-atom interaction and suppress the excess noise. The input signal mode $%
\hat{A}(t)^{in}$ is coupled into the cavity mode $\hat{a}$ through the
input-output mirror with the coupling rate to the cavity of input mode $%
\gamma _{1}=T/(2\tau )$, where T is the transmission of input-output mirror
and $\tau $ is the round-trip time of the optical mode inside optical
cavity. The other three cavity mirrors are highly reflective for the optical
signal mode, and one of them is mounted on piezoelectric transducer for
scanning or locking the cavity length. The cavity loss L is unavoidable in
real experiment due to the imperfect coating, and the corresponding decay
rate of cavity loss is $\gamma _{2}=L/(2\tau )$, which introduces the vacuum
noise $\hat{A}(t)_{\upsilon }^{in}$. The atomic spin wave decoherence rate
is $\gamma _{0}$, and couples the noise of atomic medium $\hat{S}%
(t)_{\upsilon }$ into cavity mode $\hat{a}$. When the input signal mode
resonates with cavity mode and the control mode is near-resonance with the
cavity mode, the FWM noise is off-resonance and effectively suppressed.
Quantum Langevin equations, describing evolution of observable operators for
the cavity mode $\hat{a}(t)$ and collective atomic spin wave $\hat{S}(t)$
are shown as

\begin{eqnarray}
\frac{d\hat{a}(t)}{dt} &=&-\gamma \hat{a}(t)-i\kappa (t)\hat{S}(t)+\sqrt{%
2\gamma _{1}}\hat{A}(t)^{in}+\sqrt{2\gamma _{2}}\hat{A}(t)_{\upsilon }^{in},
\\
&&\frac{d\hat{S}(t)}{dt}=-\gamma _{0}\hat{S}(t)-i\kappa (t)\hat{a}(t)+\sqrt{%
2\gamma _{0}}\hat{S}(t)_{\upsilon },
\end{eqnarray}%
where $\gamma =\gamma _{1}+\gamma _{2}$ corresponds the sum of the coupling
rate and the decay rate of cavity. For an input signal mode to be stored, a
complete mode expansion into the longitudinal modes of the input optical
mode is expressed by $\hat{A}(t)^{in}=u(t)_{0}^{in}\hat{a}_{0}^{in}$, where $%
\hat{a}_{0}^{in}$ is an optical mode operator and $u(t)_{0}^{in}$ is a
temporal mode function of the input optical mode, which determines the
optical mode shape. The input mode is dynamically shaped in time to provide
optimum memory efficiency, and the temporal mode function in our system is
approximately described by a rising exponential function \cite{39}. In the
cavity-enhanced memory system, the memory efficiency is defined as the ratio
of the photon number of the released signal mode to that of the input signal
mode, which depends on storage mechanism, media property and systematic
losses. By
solving quantum Langevin equations with the proper input temporal mode
function, the memory efficiency $\eta (T_{0})$ at the storage time $T_{0}$
from input optical mode to released optical mode is given by \cite{40}

\begin{equation}
\eta (T_{0})=\frac{(-\gamma _{1}\gamma _{0}^{2}e^{-\gamma T_{0}}+\gamma
_{1}\kappa ^{2}e^{-\gamma _{0}T_{0}})^{2}}{(\gamma _{0}+\gamma )^{2}(\kappa
^{2}+\gamma _{0}\gamma )^{2}}.
\end{equation}

\textbf{Experimental realization of cavity-enhanced quantum memory.} Fig. 2 (a) shows the experimentally measured photon fluxes of the
input and released signal modes at the storage time of 100 ns, when the
atomic cell is heated to around 95%
${}^{\circ}{\rm C}$%
. The red line and the blue line indicate the photon fluxes of the input
signal mode and the released signal mode from the cavity-enhanced quantum
memory system, respectively. The photon fluxes of the signal mode released
from the memory system after passing through a filter system consisting of
the polarizer and etalons with the external transmission of 85.6\% is
measured, thus the memory efficiency of 67$\pm $1\% is directly measured and
real memory efficiency deducting the external transmission losses should be
78$\pm $1\%.

In the quantum memory, the fidelity $F=\{Tr[(\hat{\rho}_{1}^{1/2}\hat{\rho}%
_{2}\hat{\rho}_{1}^{1/2})^{1/2}]\}^{2}$, which describes the overlap of
input states $\hat{\rho}_{1}$ and the states $\hat{\rho}_{2}$ released from
the memory system, quantifies the performance of a quantum memory.
Generally, if the average fidelities for a set of input coherent states
within a Gaussian distribution surpass the classical benchmark fidelities,
the quantum property outperforming classical systems can be verified \cite%
{41,42}. Fig. 2 (b) presents the dependence of average fidelity on mean
photon number $\overline{n}$ of the Gaussian distribution of input set of
coherent states. Blue solid line represents the theoretical average
fidelity, and the red dots are the corresponding experimentally measured
results. The black dashed line is the corresponding classical benchmark
fidelity. It shows that the average fidelity depends on the mean photon
number of the Gaussian distribution of the input coherent state, and reaches
0.97$\pm $0.01\ for the input coherent state with the same mean photon
number 0.60 of the Gaussian distribution of the input set of states, where
the corresponding benchmark fidelity is 0.73. We can see that for a set of
coherent states within the mean photon number range from $\overline{n}=0$\
to $\overline{n}=8.0$,\ the average fidelities for each input state exceed
its classical benchmark fidelity, thus the quantum property of the memory
outperforming any classical memory is confirmed \cite{35,36}\ (see
Supplementary Note 5 for details).

The memory efficiency and the average fidelity as functions of storage time
are shown in Fig. 2 (c), where the average fidelity is determined by storing
and releasing of various input coherent states with $\overline{n}=0.60$.
Blue solid line and blue squares are the theoretical memory efficiency and
the experimentally measured values, respectively; and red dotted line and
red dots represent the theoretical average fidelity and experimentally
measured results, respectively; the black dashed line stands for the
classical benchmark fidelity. From blue solid line, the lifetime in the
memory system with warm atomic cell of 1.2\ $\mu $s\ is obtained, which is
mainly limited by the magnetic noise. We can see that all average fidelities
are higher than the classical benchmark fidelities\ within the lifetime of
atoms, and the storage time can be chosen arbitrarily within the lifetime.

\begin{figure}[tbph]
\begin{center}
\includegraphics[width=16cm]{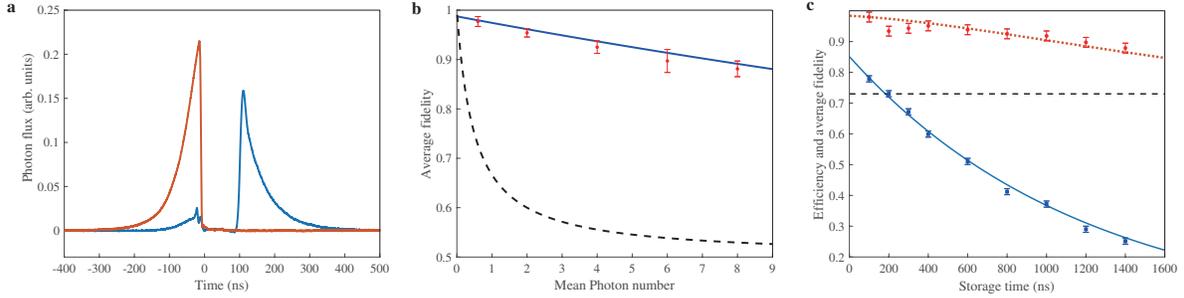}
\end{center}
\caption{\textbf{Experimental results.} \textbf{a} Temporal variation of the
experimentally measured photon fluxes. The red line and the blue line
indicate the photon fluxes of the input signal mode and the released signal
mode from the cavity-enhanced quantum memory system, respectively. \textbf{b}
The dependence of average fidelity on the mean photon number of the Gaussian
distribution set of input coherent states. Blue solid line represents the
theoretical average fidelity, and the red dots are the corresponding
experimentally measured results. The black dashed line is the corresponding
classical benchmark fidelity. \textbf{c} The results for memory efficiency
and average fidelity v.s. storage time. Blue solid line and blue squares are
the theoretical memory efficiency and the experimentally measured values,
respectively; and red dotted line and red dots represent the theoretical
average fidelity and experimentally measured results, respectively; the
black dashed line stands for the classical benchmark fidelity. Error bars
represent $\pm $1 standard error are obtained with the statistics of the
measured photon numbers and noises.}
\end{figure}

\section*{Discussion}

Due to the use of an optical cavity with the near-perfect temporal and
spatial matching, the memory efficiency of 67$\pm $1\% and the excess noise
close to QNL have been directly measured. For set of input coherent states
within the mean photon number range from $\overline{n}=0$ to $\overline{n}%
=8.0$, the deterministic average fidelities have exceeded the benchmark
fidelities. Thus the memory has entered into the quantum region.\ Because
the atomic cell is put in the single layer magnetic field shielding barrel
in the present experiment, the memory lifetime is short at the scale of
microseconds due to the influence of the residual magnetic field noise. If
the magnetic field noise is further reduced by employing the multiple-layer
structure, the lifetime must be obviously increased \cite{43}.
Alternatively, it has been demonstrated that the cell-wall anti-relaxation
coating onto the inner surface of the cell may provide an effective approach
to extend the memory lifetime of warm atom to the scale of milliseconds \cite%
{5,44}. We believe that if above mentioned feasible techniques are applied
in our system, a quantum memory with longer memory lifetime will be possibly
demonstrated based on the cavity-enhanced warm atomic system. The presented
approach is achievable on a variety of other physical platforms, such as in
trapped ions \cite{45,46,47,48}, superconductors \cite{49,50,51}, solid
states \cite{52,53,54,55} and optomechanics \cite{56,57,58,59,60}.

For optical continuous-variable (CV) quantum information systems, the
quantum information is encoded in the quadrature amplitudes and phases of
optical signal modes, which have been used to implement quantum information
protocols, such as quantum teleportation \cite{61}, quantum dense coding 
\cite{62} and quantum dense metrology \cite{63}. For the presented memory
experiment of optical coherent state, the quantum information is encoded and
stored in the quadrature amplitudes and phases of optical signal modes.
Besides, photons carrying information in discrete variable (DV), such as its
arbitrary polarization state, can also be stored in the presented system by
replacing the Glan-Thompson polarizer with a polarization insensitive beam
splitter \cite{64}. Thus, the cavity-enhanced memory works equally well for
both CV and DV quantum information because the cavity-enhanced quantum
memory is a linear mapping technique \cite{65}. Quantized states of light,
such as squeezing \cite{66} and entanglement \cite{67} are kernel resources
in quantum information science and technology, and the cavity-enhanced
quantum memory system performs well enough to preserve quantum information
of arbitrary quantized optical states. Due to the experimental simplicity of
a warm atomic cell setup \cite{68,69}, the presented system is robust and
easily controlled, which is ready to be applied in some quantum information
systems.

\section*{Methods}

\textbf{Experimental setup.} The experimental setup for implementing the
cavity-enhanced quantum memory is shown in Fig. 1 (b), and the experimental
details are given (see Supplementary Note 1). A $^{87}Rb$ atomic cell coated
with 795 nm anti-reflection, which is placed in a magnetic shielding, is
used as EIT medium of cavity-enhanced quantum memory system. An input signal
mode with an optimized wave packet originally coming from Ti:sapphire laser
is stored in an atomic cell inside the optical cavity. The time sequence of
the cavity-enhanced quantum memory is given (see Supplementary Note 2). In
this experiment, the photoreceiver detector D1 is applied to measure memory
efficiency, which is theoretically analyzed in details (see Supplementary
Note 3), and the excess noises can be analyzed from the BHD measurement (see
Supplementary Note 4).

\section*{Data Availability}

The data that support the findings of this study are available within the
paper and its Supplementary Information. Additional data are available from
the corresponding authors upon reasonable request.

\section*{Acknowledgements}

We thank Luis A. Orozco for the helpful discussion. The work was supported
by the National Natural Science Foundation of China (Grants No. 62122044
(Z. Y.), No. 61925503 (X. J.), No. 11904218 (X. J.), No. 12147215
(X. J.), No. 61775127 (Z. Y.) and No. 11834010 (Z. Y.)), the Key
Project of the National Key R\&D program of China (Grant No. 2016YFA0301402
(X. J., Z. Y., C. X., K. P.)), the Program for the Innovative
Talents of Higher Education Institutions of Shanxi (X. J.), the Program
for the Outstanding Innovative Teams of Higher Learning Institutions of
Shanxi (Z. Y.) and the fund for Shanxi \textquotedblleft 1331
Project\textquotedblright\ Key Subjects Construction (X. J., Z. Y.,
C. X., K. P.).

\section*{Author Contributions}

Z. Y., X. J. and C. X. conceived the original idea. L. M., X. L., Z. Y., X. J. and K. P. designed the experiment. L. M., X. L., J. Y., R. L. and T. C. constructed and performed the experiment.
L. M., X. L., Z. Y. and X. J. accomplished theoretical calculation
and the data analysis. Z. Y., X. J., C. X. and K. P. wrote the
paper. All the authors reviewed the manuscript.

\section*{Competing Interests}
The authors declare no competing financial interests.

\end{document}